\documentclass[aps,prd, twocolumn,longbibliography, nofootinbib, floatfix]{revtex4-1}
\usepackage{amsmath}
\usepackage{graphicx}
\usepackage{dcolumn}
\usepackage{bm}
\usepackage{epsfig}
\usepackage{amssymb,latexsym,mathrsfs}
\usepackage{mathtools}
\usepackage{commath}
\usepackage{multirow}
\usepackage{color}
\usepackage{aas_macros}
\usepackage{hyperref}
\usepackage{ragged2e}
\hypersetup{
    colorlinks=true,
    linkcolor=red,
    citecolor=blue,
} 

\newcommand{\be}{\begin{equation}}
\newcommand{\ee}{\end{equation}}
\newcommand{\bs}{\begin{split}} 
\newcommand{\bea}{\begin{eqnarray}}
\newcommand{\eea}{\end{eqnarray}}

\newcommand{\bfig}{\begin{figure}[htb]}
\newcommand{\efig}{\end{figure}}
\newcommand{\bfigc}{\begin{figure*}[htb]}
\newcommand{\efigc}{\end{figure*}}
\newcommand{\bsfig}{\begin{subfigure}[b]}
\newcommand{\esfig}{\end{subfigure}}
\newcommand{\fig}[1]{Figure~{\ref{fig:#1}}}

\newcommand{\aave}[1]{\langle #1 \rangle}

\newcommand{\omk}{\Omega_k}

\newcommand{\sdt}{\sigma_{D_{\Delta t}}}
\newcommand{\dt}{D_{\Delta t}}
\newcommand{\srs}{\sigma_{r_s}}
\newcommand{\srl}{\sigma_{r_l}}

\newcommand{\sk}{\sigma_K}
\newcommand{\somk}{\sigma_{\Omega_k}}

\begin{document} 
\title{Cosmic Curvature Tested Directly from Observations}

\author{Mikhail Denissenya${}^1$, Eric V.\ Linder${}^{1,2}$, Arman Shafieloo${}^{3,4}$} 
\affiliation{${}^1$Energetic Cosmos Laboratory, Nazarbayev University, Astana, 010000, Kazakhstan\\ 
${}^2$Berkeley Center for Cosmological Physics \& Berkeley Lab, 
University of California, Berkeley, CA 94720, USA\\ 
${}^3$Korea Astronomy and Space Science Institute, Daejeon, Korea\\
${}^4$University of Science and Technology, Daejeon, Korea}

\begin{abstract}
Cosmic spatial curvature is a fundamental geometric quantity of the 
Universe. We investigate a model independent,  
geometric approach to measure spatial curvature directly from 
observations, without any derivatives of data. This employs strong 
lensing time delays and supernova distance measurements to measure the 
curvature itself, rather than just testing consistency with 
flatness. We define two curvature estimators, with differing error propagation 
characteristics, that can crosscheck each other, and also show how they can be used to map 
the curvature in redshift 
slices, to test constancy of curvature as required by the Robertson-Walker metric. 
Simulating realizations of redshift distributions and distance measurements of lenses and 
sources, we estimate uncertainties on the curvature 
enabled by next generation measurements. The results indicate that the model independent 
methods, using only geometry without assuming forms for the energy density constituents, 
can determine the curvature at the $\sim6\times10^{-3}$ level. 
\end{abstract} 

\date{\today} 

\maketitle

%%%%%%%%%%%%%%%%%%%%%%%%%%%%%%%%%%%%%%%%%%%%%%%%%%%%%%%%%%%%%%%%%%%%%%%%
\section{Introduction} 

Spatial curvature is one of the two fundamental quantities defining 
the Robertson-Walker (RW) metric of a homogeneous and isotropic spacetime. 
Unlike spacetime curvature, which depends on the second derivative 
of the scale factor $a(t)$ entering the metric, spatial curvature 
$k$ is a purely geometric quantity. The most popular theories of 
early universe inflation predict spatial flatness $k=0$, so any 
detection of nonzero spatial curvature (hereafter just called 
curvature) would have significant impact on our understanding 
of the early universe and cosmic evolution. Moreover, if measurements 
of $k$ from observations at different redshift (or in different 
directions) significantly disagreed with each other, then this could 
call into doubt the Robertson-Walker metric and cosmic homogeneity and 
isotropy. 

Many consistency tests or``alarms'' for spatial flatness have been proposed (e.g.\ \cite{Clarkson:2007pz,
2010PhRvD..81h3537S,2011arXiv1102.4485M,
Sapone:2014nna,Rasanen:2014mca,
LHuillier:2016mtc}),
but here we focus on methods that actually deliver estimates of the 
curvature parameter $k$ or $\omk=-k/(a_0H_0)^2$, where $a_0H_0$ 
is the present expansion rate (see, for example, \cite{Bernstein:2005en,Knox:2005hx}). 
Furthermore, we aim to have the curvature derived 
directly from the observations, without any derivatives taken of 
noisy data. Finally, we want to proceed in as model independent manner 
as possible, without using any dynamics from the Friedmann equations. 
Recall that spatial curvature is a geometric quantity, and so we can in 
principle test it by purely kinematic means, without imposing any 
equations of motion. That is, we never need to know the expansion factor $a(t)$ or the 
Hubble parameter $H(a)$. 

These three principles ensure that any signals found of nonzero 
flatness, and in particular its evolution, arise from fundamental 
origins and not simply a misestimate of the matter density or numerical 
inaccuracy of differentiation of noisy data, say. Rather they will 
be as pure tests as we can enable of spatial curvature and of 
homogeneity and isotropy. 
For some other approaches to determining curvature see 
\cite{Takada:2015mma,Leonard:2016evk,Sakr:2017tsa,Witzemann:2017lhi}. 

In Sec.~\ref{sec:background} we review the relation of curvature to 
distance measurements and derive an expression for $\omk$ directly 
in terms of observables. In addition we derive a redshift dependent curvature function, 
the K test, that must hold for the RW metric. We model distance uncertainties and 
carry out their error propagation to curvature in Sec.~\ref{sec:sim}. In Sec.~\ref{sec:res} 
we discuss observational constraints from various realizations of future survey data. We 
summarize and conclude in Sec.~\ref{sec:concl}.

%%%%%%%%%%%%%%%%%%%%%%%%%%%%%%%%%%%%%%% 
\section{Curvature and Distances} \label{sec:background} 

Triangulating a surface to measure its topography has an 
exceedingly long history. The generalization to nonEuclidean spaces 
showed that angle deficits and area deficits had an intimate relation 
to curvature. However, a single triangle generally requires measurement of 
angles as well as distances to test curvature, and this is not 
necessarily practical for cosmological observations (except for 
the angle at the observer). 

Weinberg \cite{Weinberg:1972kfs} applied a volume measure formed 
from distances (distances only, no angles) between four points in Tolkien's Middle Earth, 
where a flat surface would have zero 
volume in the three space. This demonstrates a test of flatness 
(of a 2D surface) and a measurement of curvature. (In fact, for 
the distances given Middle Earth is not flat but has positive 
curvature with a measurable radius of curvature.) 

The same can be applied in cosmology, where distances between two 
points can be measured either directly, if one point is at the 
observer, or through gravitational lensing if the two points lie 
along the same light ray reaching the observer. Unfortunately, 
``cross'' distances, i.e.\ those between two lines of sight, cannot 
be measured geometrically easily (though they can statistically), and so the 
Middle Earth analogy fails. However, the expansion of the universe 
brings additional information so that a single triangle,  
associated with lens and source redshifts such that the light 
reaches the observer, can in fact 
measure curvature. 

By measuring the distance to a gravitationally lensed source, $r_s$, 
to the object doing the lensing, $r_l$, and the distance between the 
two along the null geodesic followed by the light ray, $r_{ls}$, the 
curvature can be measured. In particular, 
\be 
r_{ls}=r_s\sqrt{1+\omk r_l^2}-r_l\sqrt{1+\omk r_s^2} \ ,  \label{eq:sum} 
\ee 
where all $r$ are conformal distances. 
This follows from the properties of null geodesics in Robertson-Walker 
spacetime and does not depend at all on the Friedmann equations. 

The quantity $\omk$ is what we seek to determine. Here  
\be 
\omk=\frac{-k}{a_0^2 H_0^2} \ , 
\ee 
where $H_0$ is the Hubble constant and $a_0$ is the present 
scale factor of the universe. Recall that one cannot define 
$k=\pm1$ and $a_0=1$ simultaneously -- one can only normalize 
one or the other. 

The distances $r_l$ and $r_s$ relative to the observer can be 
measured through geometric probes such as Type Ia supernova distances 
or baryon acoustic oscillations (BAO). The distance $r_{ls}$ between 
lens and source can be found through strong gravitational lensing; here 
we focus on the use of time delays between multiple images from a 
variable source as it is closer to a geometric probe. One 
could also use measurements of the image separation, and hence 
Einstein radius, though this may be more sensitive to lens modeling 
and dynamical measurements of the lens velocity dispersion.

The time delay distance is here defined as 
\be 
\dt=\frac{r_l r_s}{r_{ls}} \ , \label{eq:ddt} 
\ee 
where we omit the conventional prefactor of $1+z_l$ since that can 
itself be directly and accurately measured. 

Putting together Eqs.~(\ref{eq:sum}) and (\ref{eq:ddt}) we can solve for  
the curvature in terms of 
observables\footnote{Ref.~\cite{Rasanen:2014mca} 
gives an equivalent expression, 
but not strictly in terms of observables.}, 
\bea
\omk&=&\frac{\dt^2}{4}\left(\frac{1}{r_l^2}-\frac{1}{r_s^2}-\frac{1}{\dt^2}\right)^2-\frac{1}{r_s^2} \nonumber\\ 
&=&\frac{1}{4}\,\left[\frac{1}{\dt^2}-2\left(\frac{1}{r_l^2}+\frac{1}{r_s^2}\right)+\dt^2\left(\frac{1}{r_l^2}-\frac{1}{r_s^2}\right)^2\right] 
\label{eq:omk}
\eea  
This will be the central equation in our analysis. 

Note that we have written all distances in terms of the dimensionless 
quantities $r_i$. For supernova and BAO distances, this is not 
unreasonable as they are measured relative to low and high redshift 
anchors respectively. Time delay distances however are dimensional 
quantities. If desired we can instead write all distance as 
dimensionful, i.e.\ $d_i=H_0^{-1}r_i$, and then the left hand side quantity of Eq.~(\ref{eq:omk}) we determine is really $\omk H_0^2$, 
or more familiarly $\omk h^2$ where $h$ is the reduced Hubble constant. 

Motivated by a first order expansion of Eq.~(\ref{eq:sum}) for 
small curvature (cf.\ \cite{Bernstein:2005en}) we could also  establish a ``K test''. 
This appears as 
\bea 
K(z_l,z_s)&\equiv& \frac{1}{\dt}-\frac{1}{r_l}+\frac{1}{r_s}\\ 
&\approx&-\frac{1}{2}\omk (r_s-r_l) + {\mathcal O}(\omk^2) \nonumber\ . 
\label{eq:Kz}
\eea  
We emphasize that in this article we use the full form of the first line to test 
curvature, not just the first order expansion. 
The K test is useful to check: 1) Is this combination of distances consistent with 
$\omk=0$?, 
and 2) If not, is its redshift dependence consistent with the  
Robertson-Walker prediction of Eq.~(\ref{eq:Kz})? We will investigate the use of both the 
full expression for $\omk$ and the K test.

%%%%%%%%%%%%%%%%%%%%%%%%%%%%%%%%%
\section{Estimating Curvature} \label{sec:sim} 

To carry out the curvature estimation we need a measurement of $\dt$ from 
a strongly lensed time delay system, and reconstructed distances at 
the redshifts $z_l$ and $z_s$, such as from a suite of supernova or 
BAO distances covering these redshifts. It would be more effective if 
the exact $r_l$ and $r_s$ could be measured directly but one is unlikely to have 
distances at exactly the right redshifts, and so must use an 
error-controlled interpolation procedure from data (using standardized 
candles -- Type Ia supernovae -- or rulers -- BAO) at neighboring 
redshifts. Fortunately there have been already several successful statistical approaches 
proposed to reconstruct the distance-redshift relation (or indeed expansion history of 
the universe) in a model independent manner,  
which can be used to estimate cosmic distances at any given intermediate redshift without 
assuming a cosmological 
model, e.g.~\cite{Shafieloo:2005nd,Shafieloo:2007cs,2010PhRvD..82j3502H,2011PhRvD..84h3501H,2012JCAP...08..002S,2012PhRvD..85l3530S}. 
Furthermore, such distances in future surveys will be much more densely measured in 
redshift than current data, simplifying the process. 

One could possibly use a 
different combination of distances from the same strong lens system 
to get $r_l$, say, 
e.g.\ from the image angular separation or Einstein radius, but this 
would introduce lens modeling uncertainties. 
(Double source plane lenses \cite{Collett:2012wv,Collett:2014ola,Linder:2016bso} offer another 
method to get geometric distances, but only as ratios of ratios.) 
We feel that systematic 
uncertainties from interpolation procedures are better understood 
than from the necessary lens modeling and dynamics. However, see 
\cite{2009A&A...507L..49P,Jee:2014uxa,Jee:2015yra}. For a phenomenological, 
non-kinematic use of time delay distances in testing curvature see 
\cite{Liao:2017yeq,Li:2018hyr}. 

The uncertainty on the determination of the curvature is given in 
terms of the measurement uncertainties on the observables by 
\bea
\somk^2&=&\frac{1}{4}\left[\left(\frac{1}{r_l^2}-\frac{1}{r_s^2}\right)^2\dt^2
-\frac{1}{\dt^2}\right]^2\left(\frac{\sdt}{\dt}\right)^2 \nonumber\\ 
&+&\left[\frac{1}{r_l^2}-\frac{\dt^2}{r_l^2}\left(\frac{1}{r_l^2}-\frac{1}{r_s^2}\right) \right]^2\left(\frac{\srl}{r_l}\right)^2 \nonumber\\
&+&\left[\frac{1}{r_s^2}+\frac{\dt^2}{r_s^2}\left(\frac{1}{r_l^2}-\frac{1}{r_s^2}\right) \right]^2\left(\frac{\srs}{r_s}\right)^2\ .
\label{eq:somk}
\eea 

All contribute at the same order of magnitude. Since the distances 
are all of order unity (i.e.\ $H_0^{-1}$) for cosmological lens 
systems, we expect the uncertainty on the curvature to be of order 
the quadrature sum of the measurement uncertainties. That is, at the few 
hundredths level for percent level distance estimates. 

Despite the nonlinear combination of distances that goes into the 
curvature estimation, the estimation has the advantage that 
the covariance matrix of the measurement errors should be 
mostly diagonal. This is a virtue of the combination of distance 
measurements used to derive the curvature: one does not expect errors 
from strong lensing time delays to be covariant with supernova or 
BAO distances, and distances measured at widely separated redshifts 
should be mostly uncorrelated (recall that strong lensing tends to 
favor $r_l\approx r_s/2$, so $z_l\approx 0.4$ and $z_s\approx 0.8$--1  
might be typical values to use). If we had instead used the angular 
diameter distance from the Einstein radius of the lens system itself, 
then covariances might tend to give more issues with systematic bias 
in the determination of curvature. 

One could also estimate the uncertainty in the K test. This is 
\be 
\sigma^2_K= 
\frac{1}{\dt^2}\left(\frac{\sdt}{\dt}\right)^2 
+\frac{1}{r_l^2}\left(\frac{\srl}{r_l}\right)^2 
+\frac{1}{r_s^2}\left(\frac{\srs}{r_s}\right)^2 \ . 
\label{eq:Kz_err}
\ee  
Again, we expect the uncertainty on $K$ to be of order  
the quadrature sum of the measurement uncertainties. 
Since $K(z_l,z_s)$ involves one factor of (inverse) distance rather 
than six, systematic bias should be even less of a worry than for $\omk$. 
Furthermore, note 
that some measurements, such as BAO, actually do measure the inverse 
distance rather than the distance itself. In any case, we specifically 
test for bias of both estimators due to nonlinear error propagation 
through analysis of simulated data; 
see Sec.~\ref{sec:res} and Appendix~\ref{sec:apxbias} for details.

%%%%%%%%%%%%%%%%%%%%%%%%%% 
\section{Observational Constraints}\label{sec:res} 

Given the expression for the two curvature test quantities and their uncertainties, 
we can estimate the signal to noise of a curvature measurement. 
First, let us make a rough estimate to guide our intuition. For strong 
lensing systems, one has a geometric focal length factor such that the distance to 
the source is approximately twice the distance to the lens (i.e.\ the lens is 
roughly midway between the source and observer). So we will particularly be interested 
in $r_s\approx 2r_l$ (we put this on a quantitative foundation below). 
For small curvature, this leads to $\dt\approx r_s$. Together, 
these have immediate implications for the estimation uncertainty in the curvature tests. 

From Eq.~(\ref{eq:somk}) we can show that under these conditions the fractional distance 
uncertainties contribute to the curvature uncertainty $\somk$ as 
\bea 
\somk^2&\approx& \frac{16}{\dt^4}\left(\frac{\sdt}{\dt}\right)^2 
+\frac{4}{r_l^4}\left(\frac{\srl}{r_l}\right)^2 
+\frac{16}{r_s^4}\left(\frac{\srs}{r_s}\right)^2 \label{eq:somk1}\\ 
&\approx& \frac{16}{D^4}\left[\left(\frac{\sdt}{\dt}\right)^2 +4\left(\frac{\srl}{r_l}\right)^2+\left(\frac{\srs}{r_s}\right)^2\right]\ , 
\eea 
where in the second line we use the rough approximation $D\equiv r_s\approx \dt\approx 2r_l$. (For $z_l=0.3$, $z_s=0.6$ the ratios are $\dt/r_s=1.18$, $r_s/r_l=1.85$ for 
$\omk=0$.) 
Thus, for $\omk$ determination we roughly 
care about the quadrature sum of fractional distance uncertainties, 
but the uncertainty in $r_l$ gets more weight. This is fortunate 
since we expect this distance to be the best determined. 

Comparing to the K test, from Eq.~(\ref{eq:Kz_err}) we see that $\sigma_K$ is a factor 
of 4 smaller in each of the terms, except for a factor 2 smaller in the fractional lens 
distance uncertainty. Since the lens distance uncertainty is likely to be subdominant, 
the rough expectation is that $\somk\approx(4/D)\sigma_K$. In Hubble units, $D\approx1$ 
for $z_s\approx 1$. Note that also $K\approx -\omk D/4$. 

Figure~\ref{fig:kz} shows the curvature quantities and uncertainties as a function of $\omk$ 
for a strong lens system with $z_l=0.6$ and $z_s=1.2$. We adopt a fractional time 
delay distance precision of 3\%, and lens and source distance precisions (from, e.g., 
supernova or BAO distances) of 1\%~(e.g.~\cite{Aghamousa:2016zmz}). We also verified that  
\{$\sdt$,$\srl$,$\srs$\}=\{3\%, 0.75\%,1.2\%\}, which has the same quadrature sum, gives substantially similar results.

%%%%%%%%%%%%%%%%%%%%%%%%%%%%% 
\bfig
\includegraphics[width=\columnwidth]{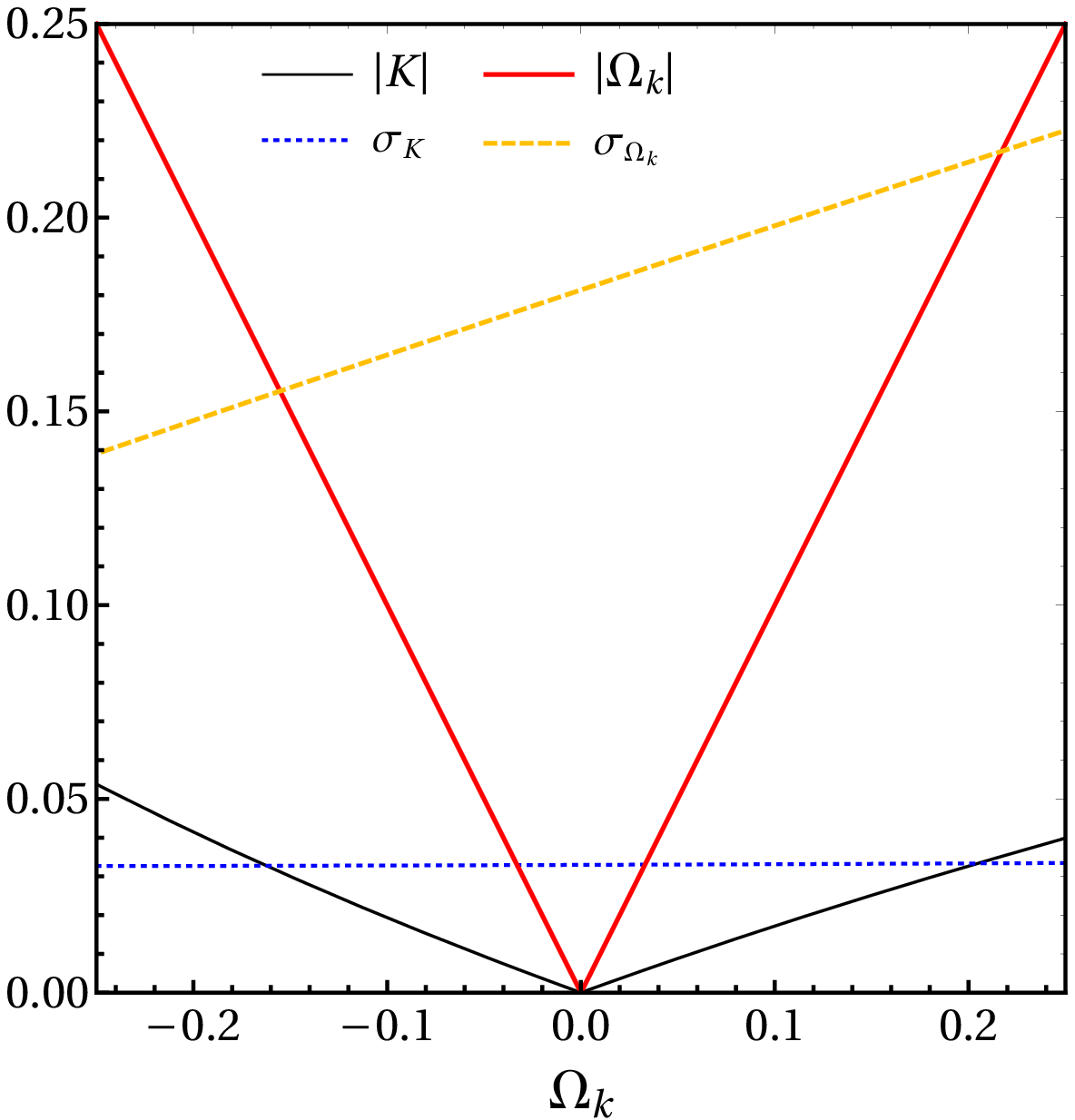}
\caption{Absolute values and uncertainties of the two curvature quantities $\omk$ and $K(z_l,z_s)$ plotted vs $\omk$. The ratio of the 
value to the uncertainty is unity where the respective curves cross. The uncertainties are  
for a single lens system at $z_l=0.6$, $z_s=1.2$ with fractional distance uncertainties 
on $\dt$, $r_l$, and $r_s$ of $\{3\%,1\%,1\%\}$ respectively. 
}
\label{fig:kz}
\efig

The intersection point between the curvature quantity ($\omk$ or $K$) curve and its 
uncertainty curve determines the lower bound on the measurement of the curvature parameter for a given strong lens system. Due to the scaling discussed above, in fact 
$\somk/|\omk|\approx \sigma_K/|K|$ (which we will abbreviate 
as the inverse signal to noise $S/N$) and so the intersections correspond to nearly the same 
value of $\omk$ that can be distinguished from flatness. For example, the $S/N=1$ for 
the $\omk$ quantity at $\omk=-0.15$ or $\omk=+0.20$, and the same holds for the 
$K$ test, for $\{3\%,1\%,1\%\}$ precision. Thus a single such system could distinguish 
$\omk<-0.15$ from $\omk=0$ at $S/N>1$. 

These results were for a single system, with fixed $z_l=0.6$ and $z_s=2z_l$ so we next 
investigate the sensitivity to the two redshifts, including the optimum, and the 
improvement enabled by large numbers of systems delivered by next generation strong 
lensing, supernova, and BAO surveys. 

Considering the redshift distribution, we expect that the raw 
sensitivity should improve for large $r_s$ (i.e.\ $D$) since this 
lowers the uncertainty at fixed $\sigma_D/D$. However, systems at high 
redshift would likely be less well constrained observationally and so $\sigma_D/D$ 
would in fact increase. The interplay with the observational accuracy 
depends on the specific type of measurement, e.g.\ BAO or supernovae, 
what type of BAO (e.g.\ galaxies, quasars, etc.), and survey 
strategy and specifics and is beyond the scope of this investigation. Instead 
we will present results fairly generally, taking as a baseline a 
conservative approach of medium redshifts ($z_l\lesssim0.6$) and discussing the 
impact of higher redshift observations if they can be accomplished 
with good accuracy (perhaps in ``golden'' systems). 

Figure~\ref{fig:sno} plots the S/N in the determination of curvature $\omk$ and the K test as 
a function of the two redshifts $z_l$ and $z_s$, for the 
case with $\omk=-0.05$. As expected from the previous discussion, the two tests are 
substantially similar. The highest $S/N$ occurs for the highest $z_s$, under the 
assumption of constant fractional precision. 
We see that the optimum for a given $z_s$ indeed occurs at $z_l\approx z_s/2$. 
(This gives $r_l$ slightly greater than $r_l\approx r_s/2$ since by 
Eq.~\ref{eq:somk1} we want somewhat higher $\dt$ 
and $r_l$ to reduce the error on $\omk$.)

%%%%%%%%%%%%%%%%%%%%%%%% 
\bfig
\includegraphics[width=\columnwidth]{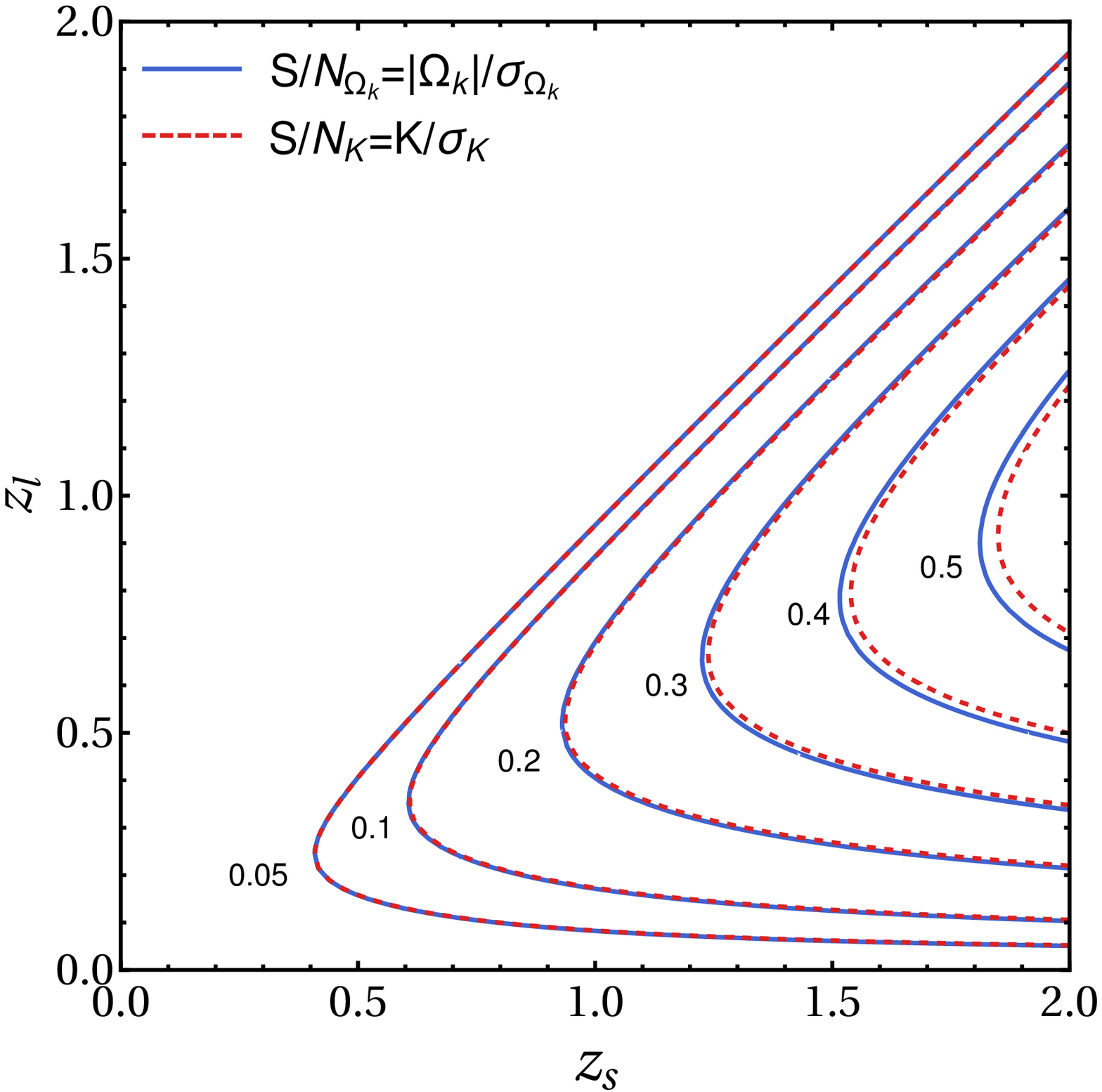}
\caption{Isocontours of $S/N_{\omk}=\abs{\omk}/\somk$ and $S/N_K=K/\sk$ are plotted in 
the $z_s$-$z_l$ plane, for fixed $\omk=-0.05$ and $\{3\%, 1\%,1\%\}$ fractional distance 
precision.} 
\label{fig:sno}
\efig

Our canonical $z_l=0.6$, $z_s=1.2$ gives a $S/N\approx0.3$, so some 
10 systems would be required to get $S/N=1$ for distinguishing 
$\omk=-0.05$ from 0. Pushing to $z_l=0.9$, $z_s=1.8$ could raise 
the $S/N$ to 0.5, but at the price of longer and more difficult 
observations to reach the same fractional distance precision. 

The redshift dependence of the uncertainty in $\omk$ and K estimations 
is presented in more detail in Fig.~\ref{fig:komkbias}, for the moment fixing $z_s=2z_l$ 
since this gives close to the optimum. 
As expected the uncertainties decrease for higher $z_l$ (and 
hence higher $z_s$, going roughly as $(1+z_l)^{-4.3}$ for 
$\somk$ and $(1+z_l)^{-2.7}$ for $K$. While $\omk$ is of course 
constant with redshift, $K$ increases so $\sk/K$ and 
$\somk/\omk$ keep nearly in step over the redshift range of 
interest. Furthermore, the statistical uncertainties decrease 
as the square root of the number of systems $n$, so the $S/N$ 
improves as $1/\sqrt{n}$. We also 
plot the systematic bias $\delta K$ and $\delta\omk$ due to 
the nonlinearity of the error propagation; these are negligible in comparison to the 
statistical uncertainties for $n<10^4$ and can be controlled further as we discuss below 
and in Appendix~\ref{sec:apxbias}.

%%%%%%%%%%%%%%%%%%%%%%%%%% 
\bfig
\centering
\includegraphics[width=\columnwidth]{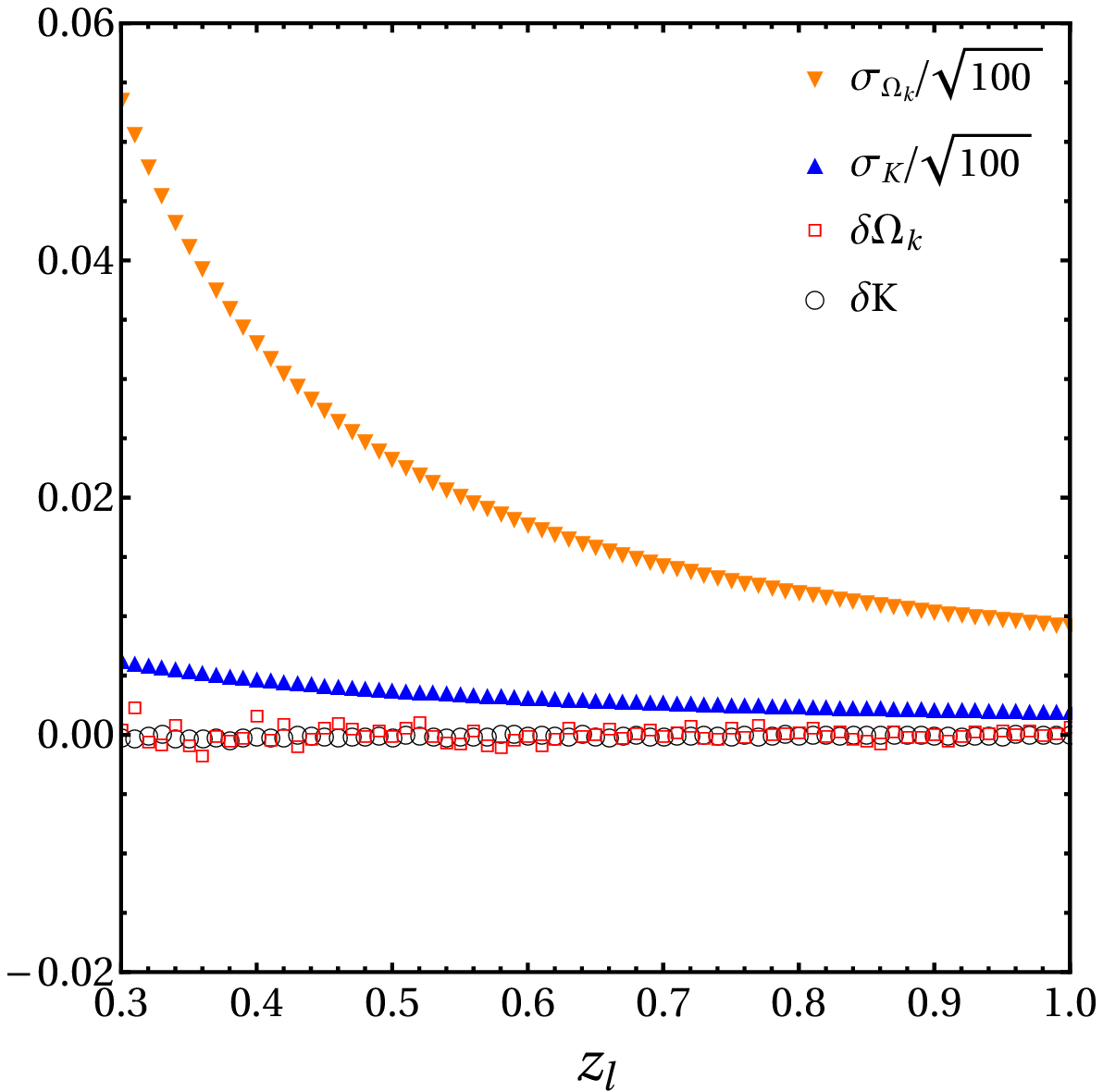}
\caption{Standard deviations $\sk/\sqrt{100}$ and $\somk/\sqrt{100}$ (corresponding to 
100 systems) and biases $\delta K=\aave{K}-K_{\text{true}}$ and 
$\delta\omk=\aave{\omk}-\Omega_{k,\text{true}}$ are plotted versus $z_l$. The averages 
are weighted as discussed in the text. 
}
\label{fig:komkbias}
\efig

Next we consider ensembles of measurements over a range of redshifts such as would be 
delivered by next generation surveys. Set A has $z_l\in[0.3,0.6]$, representing a medium depth survey, 
and Set B has $z_l\in[0.3,0.9]$, for a deep survey, both with a source distribution 
$z_s=[1.5\,z_l,2.5\,z_l]$. 
While there is more volume at higher $z_l$, the measurements are 
more difficult; we do not compute a lens redshift distribution, which would depend 
on survey specifics such as magnitude depth, cadence, etc., but rather sample $z_l$ randomly from a uniform distribution in the 
given ranges. The source redshift $z_s$ is then sampled uniformly from its corresponding range. We 
study results for 100, 400, and 800 systems, to check statistical 
scaling. 

An instantiation of the sets with 100 systems each is shown in \fig{ab}. For the distances, 
we sample from normal distributions to incorporate observational uncertainties. 
For every redshift pair of $z_l$ and $z_s$, from either Set A or B, we compute $\dt$, 
$r_l$, $r_s$ distances and multiply each by a random number drawn from the normal distribution with unit mean and standard deviation set to measurement fractional 
precisions $\{3\%, 1\%,3\%\}$ respectively. 
This gives realizations of all the simulated data, from which can then be computed the $\omk$ 
and $K$ quantities.

%%%%%%%%%%%%%%%%%%%%%%% 
\bfig
\centering
\includegraphics[width=\columnwidth]{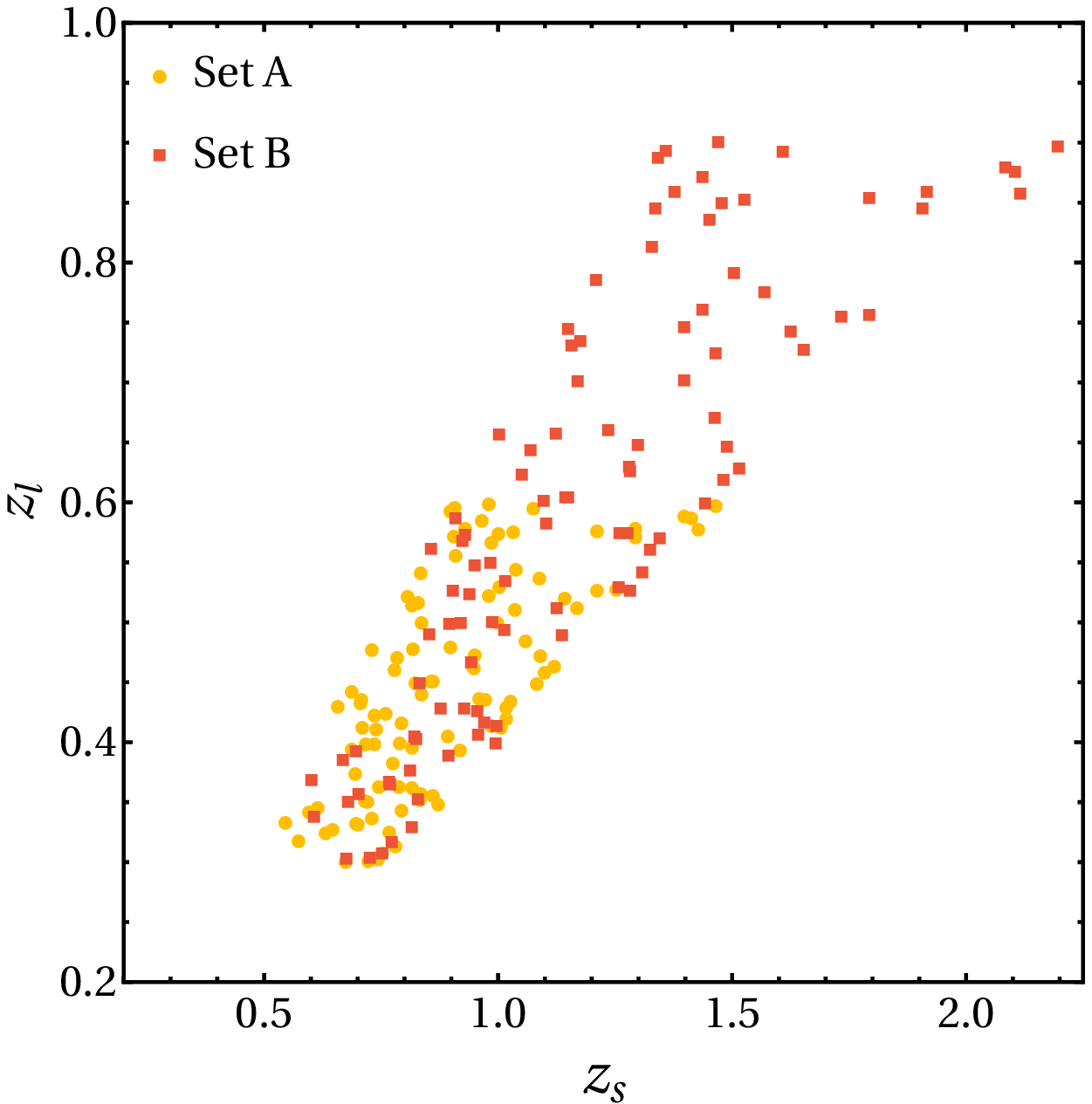}
\caption{Sets A and B of $N$=100  uniformly  distributed $(z_l,z_s)$ pairs with $z_l$ $\in$ $[0.3,0.6]$ and $z_l$ $\in$ $[0.3,0.9]$ respectively, and  $z_s=(2\pm0.5)z_l$ for each set.}
\label{fig:ab}
\efig

We are particularly interested in the discriminatory power of the estimated quantities to make 
a statistically significant detection of curvature compared to flatness. This can be 
thought of in terms of a ``signal to noise'' ratio, defined as 
$S/N_X=\aave{X}/\aave{\sigma_X}$, 
where $X$ is either $\omk$ or $K$. The ensemble standard deviation is 
propagated from independent individual measurements by 
  \be
 \aave{\sk}=\frac{1}{\sqrt{\sum \frac{1}{\sigma_{K_i}^2}}}\ . 
  \ee 

For each particular value of $\omk$, there is an 
associated $S/N_X$ derived from the ensemble of data. This allows us to investigate 
several important characteristics: 1) for what range of $\omk$ can we disfavor flatness? 
2) which test -- estimation of $\omk$ or the K test -- is more incisive?, and 
3) how does the estimation scale with data set, i.e.~redshift range, number density, and 
total number? 

Figures~\ref{fig:Akomk} and \ref{fig:Bkomk} show the results. First, note that for small 
values of $\omk$ the $S/N$ is rather linear with $|\omk|$, and fairly insensitive to 
whether $\omk<0$ or $\omk>0$. Of course, $S/N=0$ for the 
flat universe since by definition there is no signal of deviation from flatness. 
Second, the $\omk$ estimation and the K test give nearly the same results, as motivated 
earlier, so we are 
free to use either (and they do have different error propagation so consistency of 
results is a good crosscheck). Third, in the most pessimistic of our scenarios (Set A with 100 systems), the 
data achieves $S/N>1$ for $|\omk|>0.025$ and in the most optimistic of our basic 
scenarios (Set B with 400 systems), $S/N>1$ for $|\omk|>0.008$.

%%%%%%%%%%%%%%%%%%%%%% 
\bfig
\centering
\includegraphics[width=\columnwidth]{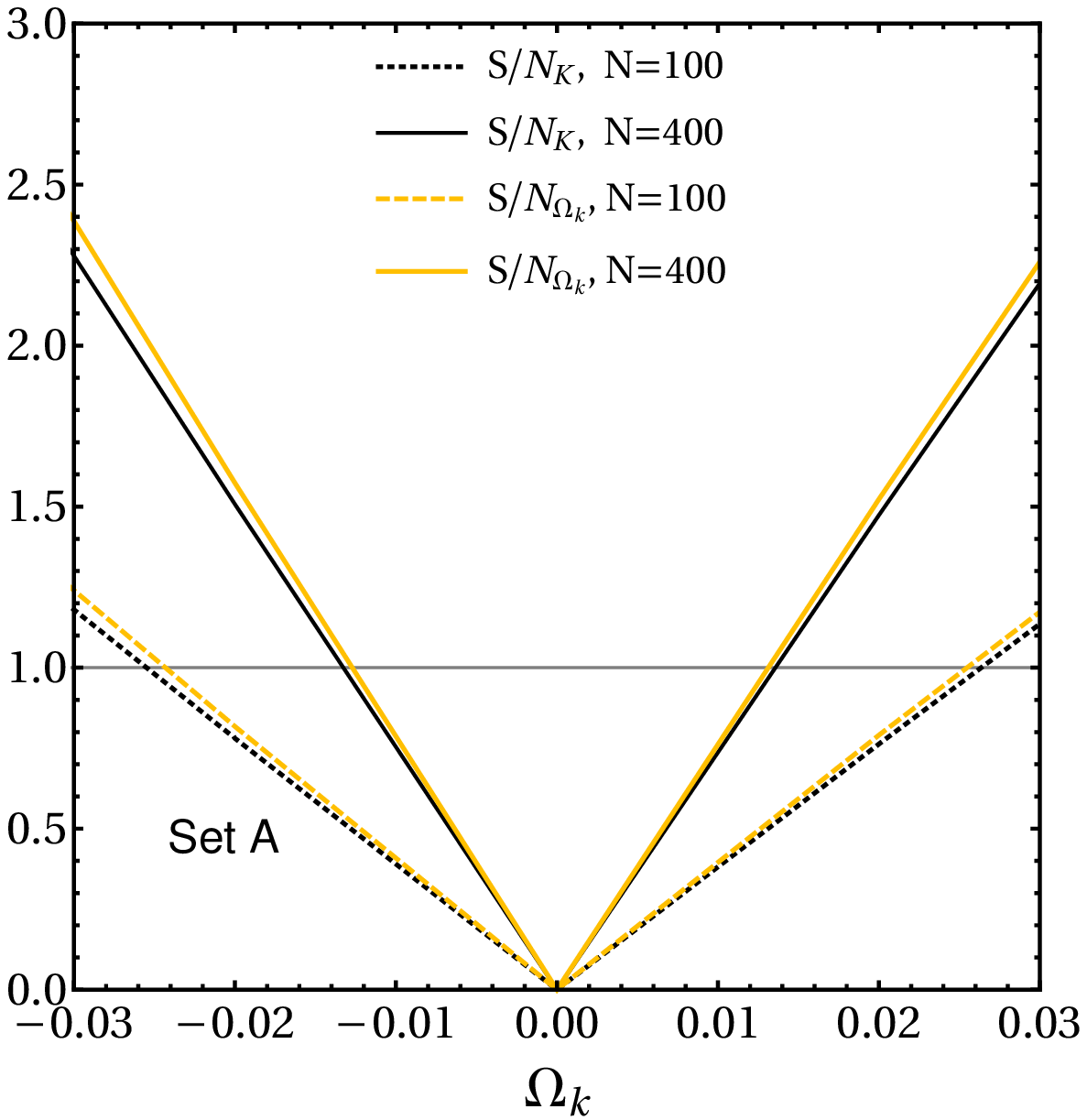}
\caption{Signal-to-noise ratios $S/N_K=\abs{\aave{K}}/\aave{\sk}$ and $S/N_{\omk}=\abs{\aave{\omk}}/\aave{\somk}$ versus curvature parameter $\omk$ for the data realization of Set A, with 100 or 
400 systems total. The light grey horizontal line indicates $S/N$=1, and hence the intersection with it gives the constraint 
on $\omk$.}
\label{fig:Akomk}
\efig

%%%%%%%%%%%%%%%%%%%%%% 
\bfig
\centering
\includegraphics[width=\columnwidth]{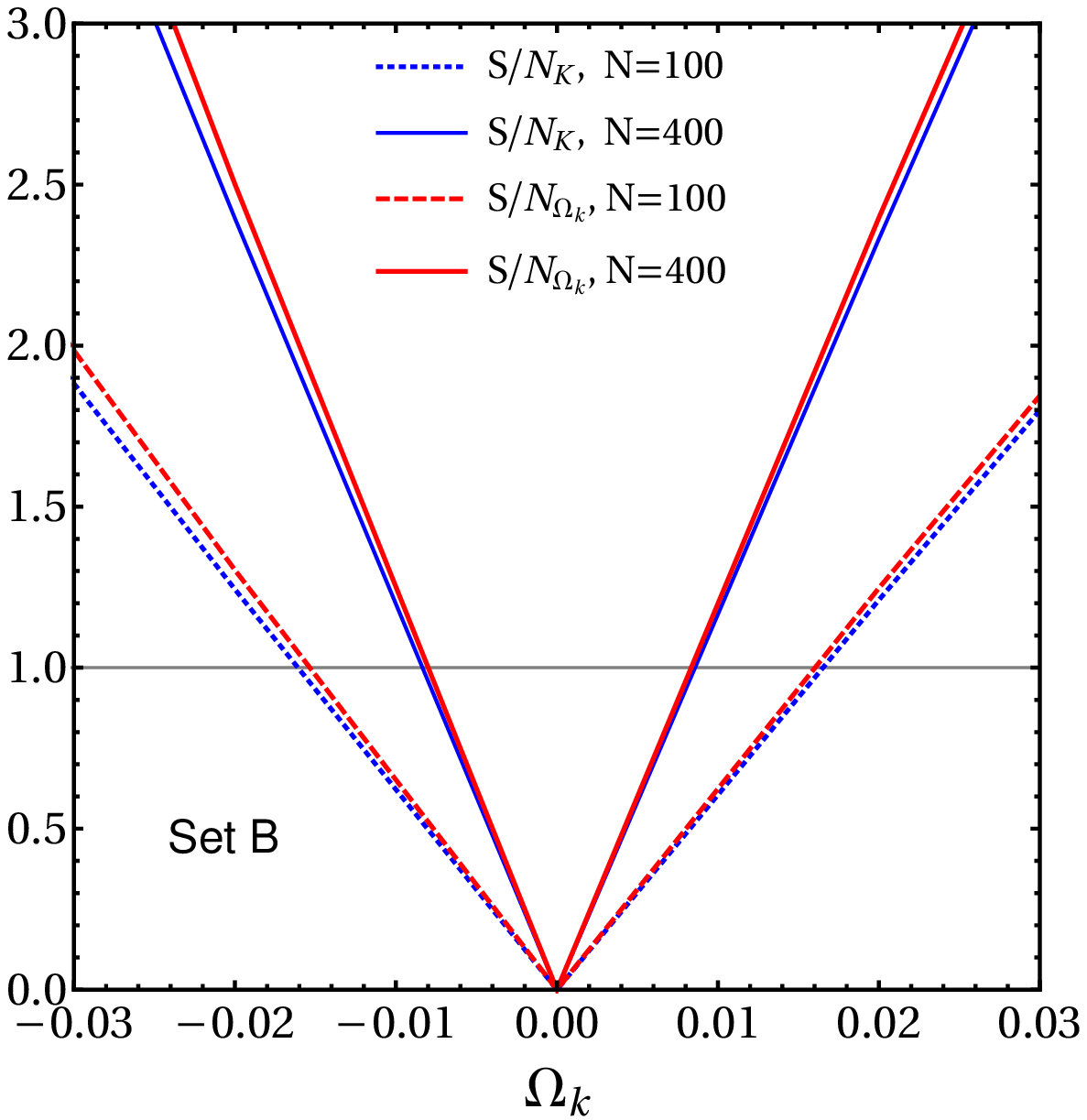}
\caption{As Fig.~\ref{fig:Akomk}, for data realization Set B.}
\label{fig:Bkomk}
\efig

As for scaling, the $S/N$ goes as the square root of the total number of 
data points, but we also see that redshift range plays a strong role. 
For the same number of systems, Set B provides a $\sim50\%$ increase over 
Set A in $S/N$ despite having half the number density of systems. More 
quantitative detail appears below.

One of the virtues of the K test is that it has a particular predicted redshift 
dependence, given by Eq.~(\ref{eq:Kz}). To explore this, we can subdivide the data 
into redshift bins, e.g.\ of width $\Delta z=0.1$. For a relatively small data 
sample this can give large scatter within a bin. While the statistical dispersion is 
simply the price paid for a modest data sample, one must also pay careful attention 
to the bias induced by nonlinearity of the error propagation from fluctuations to 
large uncertainties. We therefore introduce weighting that gives priority to measurements 
with small uncertainties and reduces the impact  
of data with large uncertainties. 

The most straightforward implementation is inverse variance weighting. We apply this 
to the K test data, since this is predominantly what we want to subdivide. Defining 
\be
 \aave{K}=\frac{\sum \frac{K_i}{\sigma_{K_i}^2}}{\sum \frac{1}{\sigma_{K_i}^2}} \ , 
\ee
where $i$ runs over each data point within the desired subsample (i.e.\ redshift bin), 
we find that this greatly ameliorates bias, keeping it much less than the 
statistical dispersion. We discuss this in detail in Appendix~\ref{sec:apxbias}. 

Figure~\ref{fig:khist} plots the individual redshift bin measurements, with error bars, for 
a data realization of the 
optimistic survey of Set B, with 800 systems total. We choose to plot 
the optimistic case so that the eye can readily discern that the 
$S/N>1$. For less optimistic cases with $S/N\approx1$, the eye 
cannot recognize the pattern in the scatter as easily. In the 
case plotted, we can see the clear trend of increased deviation 
of $K$ from zero with redshift, and that it is consistent with 
the curve expected (Eq.~\ref{eq:Kz}) from $\omk=-0.02$ (the actual input to the 
realization). In a $\Delta\chi^2$ sense, the simulated data 
discriminates from zero curvature at about $3.5\sigma$, i.e.\ 
the estimation of curvature has uncertainty $\somk\approx0.006$.

%%%%%%%%%%%%%%%%%%%%%%%%%%%
\bfig
\centering 
\includegraphics[width=\columnwidth]{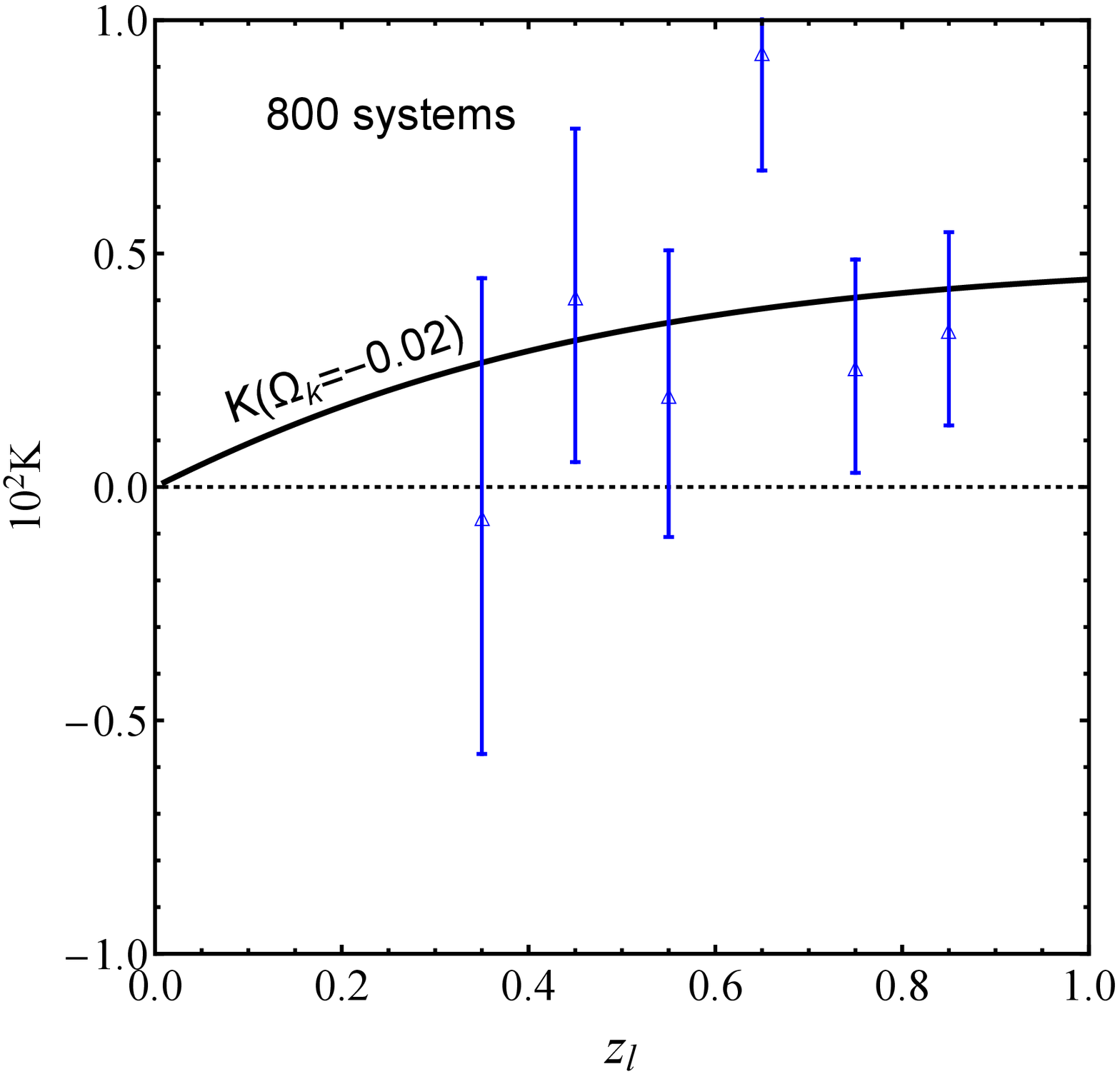}
\caption{The K test gives a specific redshift dependence in the presence 
of curvature. The solid curve is the theoretical prediction for $K(z_l,z_s)$ versus $z_l$ from Eq.~(\ref{eq:Kz}), for $\Omega_k=-0.02$. Points show the 
results of our simulated measurements $\aave{K}\pm\sigma_K$ in bins of $z_l$ for Set B with 800 systems total. 
Deviations from $K=0$ point to the existence of curvature, and the redshift dependence both estimates $\omk$ and helps separate the signal from systematics.  
} 
\label{fig:khist}
\efig

The trend for other data set realizations is fit fairly well by 
\be 
\Delta\chi^2\approx C\,(n/800)^{-1}\ , \label{eq:chiscale} 
\ee 
for discrimination of $\omk=-0.02$ from flatness by $n$ systems, where $C\approx12$ for 
Set B data (long redshift range) and $C\approx 6$ for Set A data (short redshift range). 
This corresponds to 
\be 
\somk\approx D\,(n/800)^{-1/2}\ , \label{eq:somkscale} 
\ee 
where $D\approx 0.006$ (0.009) for Set B (A) respectively. Note the extended redshift range 
(at constant fractional precision) is worth a factor $\sim2.4$ in number, 
i.e.\ Set B with 330 systems has approximately the same leverage as Set A with 800 systems. 

The curvature $\omk$ has no redshift dependence and so direct estimation from 
Eq.~(\ref{eq:omk}) does not require any subdivision with redshift. Analyzing the simulated 
data set Set B with 800 systems total (generated with $\omk=-0.02$) as a whole gives 
$\Delta\chi^2=-15$ with respect to zero curvature, a clear signature at the 
$\sim3.9\sigma$ level. For this approach, the values of $C$ and $D$ in 
Eqs.~(\ref{eq:chiscale}) and (\ref{eq:somkscale}) are 15 and 0.005 for Set B and 7.5 and 
0.007 for Set A. 

We note that $\somk$ has only a weak dependence on the value of 
$\omk$, with $\somk$ for 800 systems changing by $\sim 10^{-4}$ for a 
change in $\omk$ of 0.02 (and $\sk$ is even more insensitive).

In addition to the ensemble evaluation of curvature, we might also like to examine the 
curvature estimation with 
redshift, to assure ourselves of its constancy and check for 
systematics. Performing the same redshift binning as described 
above, and weighting the estimation of $\omk$ as described in 
Sec.~\ref{sec:apxbias} to control bias due to nonlinear error propagation, 
we present the results in Fig.~\ref{fig:omkhist}. The results have very similar 
$\Delta\chi^2$ to the unbinned case, and so again have strong discrimination for 
the true input model over zero curvature.

%%%%%%%%%%%%%%%%%%%%% 
\bfig
\centering
\includegraphics[width=\columnwidth]{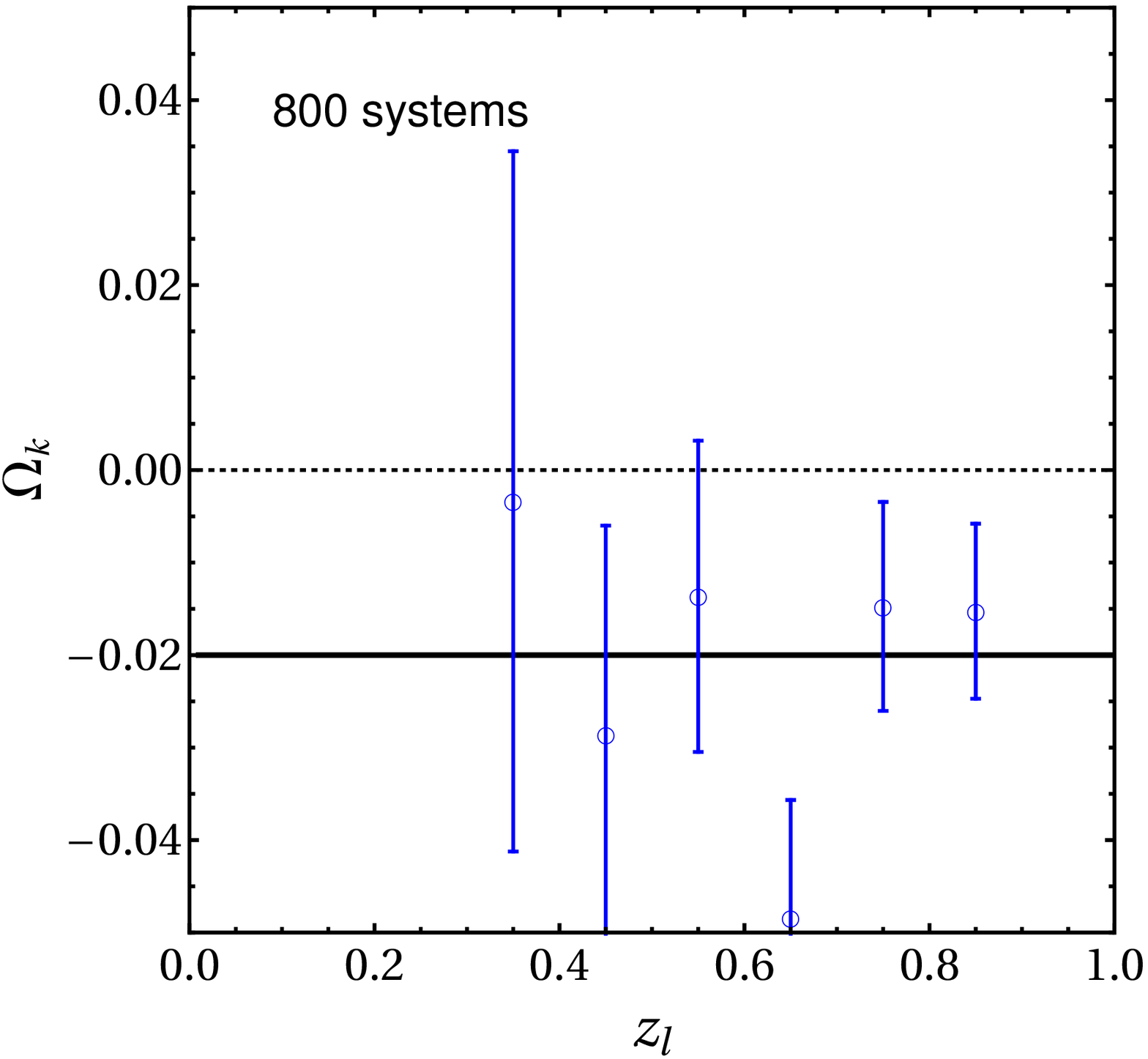}
\caption{The curvature test in redshift bins has leverage on both estimation of the value of 
the curvature, and a check on its constancy. 
The solid curve shows the input value $\omk=-0.02$. Points show the 
results of our simulated measurements $\aave{\omk}\pm\somk$ in bins of $z_l$ for Set B with 800 systems total. 
}
\label{fig:omkhist} 
\efig

%%%%%%%%%%%%%%%%%%%%%%%%%%%%%%%
\section{Conclusion}\label{sec:concl} 

We sought a robust method of estimating cosmic spatial curvature with three key 
aspects: 1) it provides an accurate estimate rather than simply a consistency test 
or alarm, 2) it derives directly from the observations, without taking derivatives 
or extrapolations of data, and 3) it is as model independent and purely geometric as 
possible. Two quantities, the direct curvature estimate $\omk$ and the K test, 
satisfied these criteria, using strong gravitational lensing time delay distances and 
supernova or BAO distances. 

Each method could provide a crosscheck of the other, with different error propagation. 
Furthermore, the K test involves a redshift dependence for the influence of curvature, 
allowing not only a fit but a verification that the proper functional dependence is 
satisfied. We examined the error propagation in some detail, demonstrating that not 
only was bias negligible compared to statistical scatter, but could be further 
reduced by appropriate weighting or Monte Carlo simulation. The two estimators give 
similar constraining power, or signal to noise, for data in the range $z_l\lesssim1$. 

Carrying out an optimization study, we found that systems with source redshift 
$z_s\approx 2z_l$ had the most leverage, close to the natural kernel for strong lensing 
systems. After an individual redshift analysis to build intuition, we then simulated 
data sets of various size and redshift range, corresponding to nearer term and next 
generation distance surveys. We presented analysis of the scaling of the curvature 
constraints with survey number, number density, and redshift range, finding that the 
range was the most important, slightly more so than number. For example,  
Set B had twice the redshift range of Set A and only needed $\sim40\%$ the number of systems to achieve the same constraining power. However, this assumed the measurement 
precision could be maintained to higher redshift. 

Figures~\ref{fig:khist} and \ref{fig:omkhist} present illustrations of how applying 
such curvature analysis to next generation surveys might appear. The K test naturally 
gives not only a signal of deviations from the zero result of a spatially flat 
universe but an accurate measurement of the curvature value and a quantitative test of 
the predicted redshift dependence. The curvature estimator can either apply to the 
data as a whole, or also divided into redshift slices, allowing a direct check of its 
predicted constancy with redshift. Again, recall the predictions are model independent, 
relying only on the Robertson-Walker metric. 

One can of course estimate the curvature parameter in a model dependent manner, relying 
on the Friedmann equations for example. This will generically give tighter constraints, 
but can be sensitive to the assumed model. For example, misestimation of the dark 
energy equation of state or the number of effective neutrino species may lead to 
biases comparable to the desired precision. For our model independent approach, 
practical application will require careful attention to nuisance parameters in the 
distance determinations, covariance of measurements, etc. 
We did not consider Hubble parameter measurements, both because of systematics 
issues and because even if model independent they are point measures rather than 
the sort of triangulation we look for in probing the spatial geometry. 

Spatial curvature is a fundamental aspect of the universe, and could hold deep clues 
to inflation and cosmic origins. Estimators formed from the symmetry properties of the 
Robertson-Walker metric in a model independent manner directly test homogeneity and 
isotropy. Such methods as we discussed here can be an important complement to other 
cosmological techniques in exploring the nature of our universe.

%%%%%%%%%%%%%%%%%%% 
\acknowledgments 

This work is supported in part by the Energetic Cosmos Laboratory and by the U.S.\ Department of Energy, Office of Science, Office of High Energy Physics, under Award DE-SC-0007867 and contract no.\ DE-AC02-05CH11231. A.S. would like to acknowledge the support of the National Research Foundation of Korea (NRF- 2016R1C1B2016478) and Energetic Cosmos Laboratory, Nazarbayev University for hospitality during the preparation of this work.

%%%%%%%%%%%%%%%%%%%%%%%%%% 
\appendix 

\section{Dealing with Bias} \label{sec:apxbias} 

Like all nonlinear functions of the measurements, formally the K test and estimate 
for $\omk$ are biased due 
to the nonlinear propagation of measurement uncertainties. However, this is of 
little real concern in the present case. For next generation data sets of less than 
1000 strong lens 
systems (i.e.\ measurements of $\dt$), the scatter dominates over the bias. 
If we look beyond this, there are two straightforward methods for dealing with 
bias: weighting and Monte Carlo simulation. 

The quantity $K$ in the K test involves inverse distances. Sometimes this is actually 
what is measured, as for BAO transverse angular scales. But if the measurement 
uncertainties are Gaussian distributed in the distance itself, with mean $D_0$ and 
standard deviation $\sigma_D$, then the mean of the inverse distance is 
$\aave{1/D}=(1/D_0)[1+(\sigma_D/D_0)^2]$. Hence there is a bias of fractional 
magnitude $(\sigma_D/D_0)^2$. 

However, we also see the way around this, by weighting 
the quantity to be averaged such that it appears more linear. This can be accomplished 
by multiplying $K\sim 1/D$ by $1/\sk^2$, where the latter is similar to $D^2/(\sigma_D/D)^2$. 
That is, we expect inverse variance weighting to ``debias'' $K$. For the $\omk$ 
estimation the situation is more complicated since there are multiple powers of 
multiple distances. For example, if the first terms in Eqs.~(\ref{eq:omk}) and 
(\ref{eq:somk}) dominate, then the $D$ portion is linearized by weighting by 
$1/\somk^{1/2}$. Other terms will prefer different weighting. In the end, while 
informed by such heuristic arguments we rely on purely empirical analysis to determine 
what weighting is most successful in debiasing $K$ and $\omk$, testing for several 
redshifts and inputs $\omk$. 

We conclude that inverse variance weighting works well for $K$ and $1/\somk^{0.35}$ 
for $\omk$. Figure~\ref{fig:biaskwt} and Figure~\ref{fig:biasowt} demonstrate the 
results. The true, input behavior is recovered to excellent approximation. Note 
for example that the leftmost $K$ point, corresponding to $z_l=0.3$, has a 
systematic bias at the $2\times10^{-4}$ level, while the statistical dispersion is 
larger than this as long as we are dealing with $n<10^5$ systems at this redshift.

%%%%%%%%%%%%%%%%%%%%%%%%%%%%%%% 
\bfig
\centering
\includegraphics[width=\columnwidth]{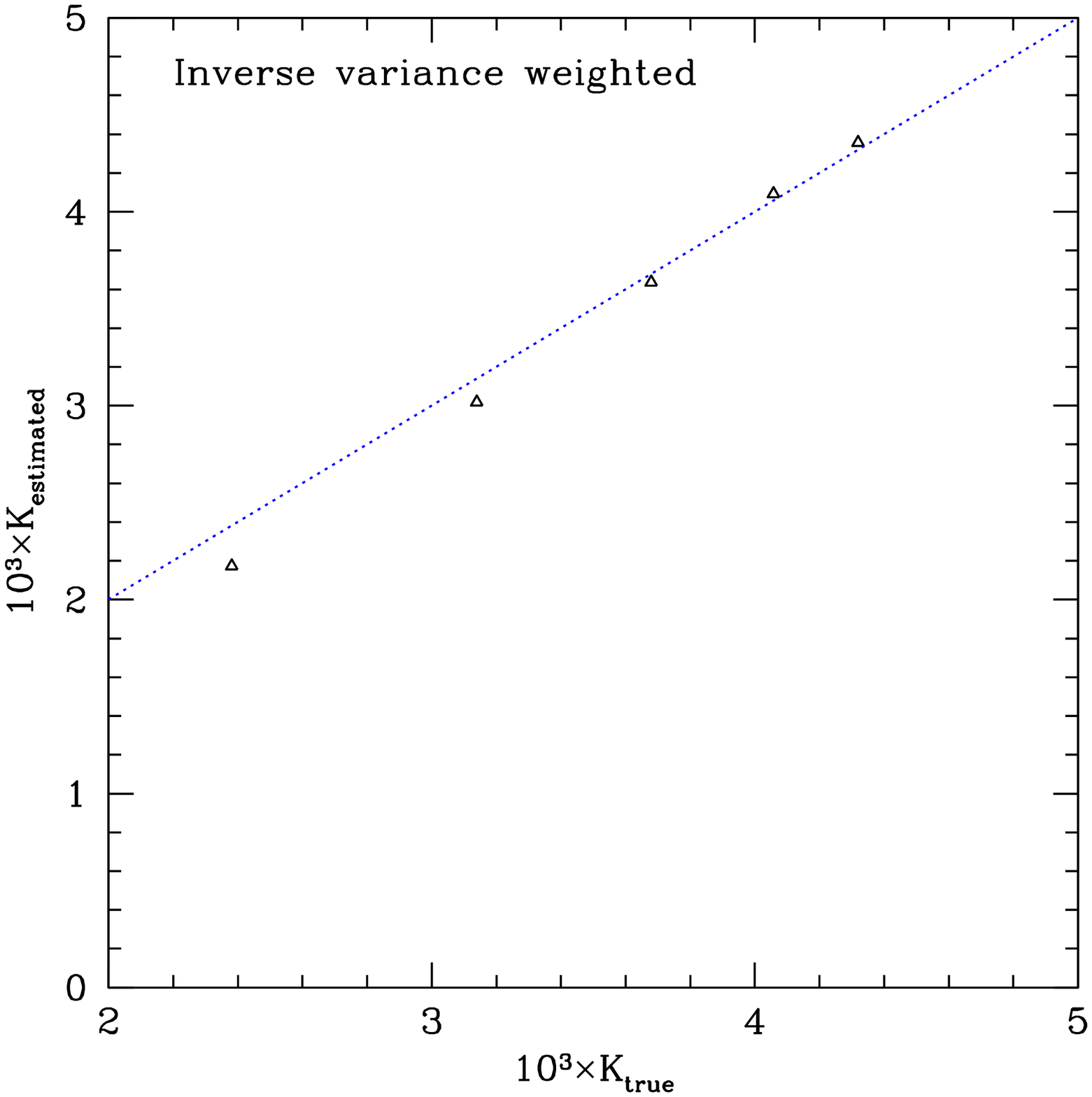}
\caption{Inverse variance weighting in the K test provides a substantially 
unbiased result, where $K_{\rm estimated}=\sum (K_i/\sigma_{K_i}^2)/\sum (1/\sigma_{K_i}^2)$. The estimates for $z_l=0.3$, 0.45, 0.6, 0.75, 0.9 
from left to right (triangular points) lie very close to $K_{\rm true}$ 
(dotted blue line), and the residual bias for this case of $\omk=-0.02$ is small compared to the statistical dispersion and the distinction from the flat case 
$K_{\rm estimated}=0$.}
\label{fig:biaskwt}
\efig

%%%%%%%%%%%%%%%%%%%%%%%%%%%%% 
\bfig
\centering
\includegraphics[width=\columnwidth]{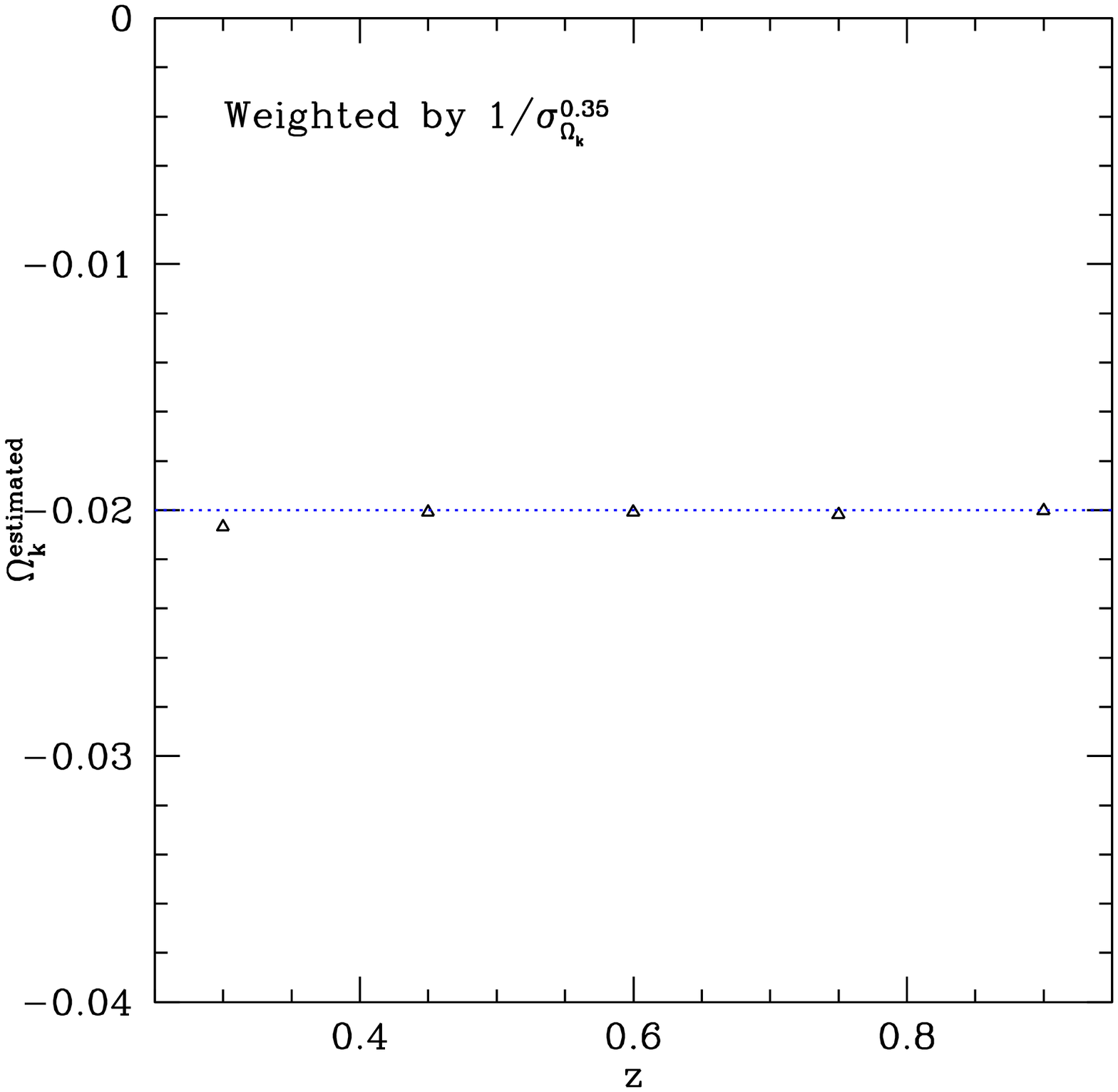}
\caption{Appropriate weighting on the $\omk$ estimator can provide a substantially 
unbiased result; we find that $\Omega_{k,\rm estimated}=\sum (\Omega_{k,i}/\sigma_{\Omega_{k,i}}^{0.35})/\sum (1/\sigma_{\Omega_{k,i}}^{0.35})$ 
works quite well. The estimates for $z_l=0.3$, 0.45, 0.6, 0.75, 0.9 
from left to right (triangular points) lie very close to the true value 
$\Omega_{k,\rm true}=-0.02$ (dotted blue line). 
}
\label{fig:biasowt}
\efig

We emphasize that the weighting, and the results in the figures, should be 
viewed simply as a demonstration of principle that bias can be reduced. The 
actual data analysis should employ Monte Carlo simulations of the actual 
data characteristics. This can furthermore be done iteratively: e.g.\ subtract the 
modeled bias for $\omk=0$, estimate the new $\omk$, resimulate, etc.

\bibliography{bbl}

\end{document}